\def \ln {{\rm \ ln \  }}

\def \a{\alpha}
\def \b{\beta}
\newcommand{\rf}[1]{(\ref{#1})}
\newcommand{\beq}{\begin{equation}}
\newcommand{\eeq}{\end{equation}}
\newcommand{\bea}{\begin{eqnarray}}
\newcommand{\eea}{\end{eqnarray}}
\newcommand{\beas}{\begin{eqnarray*}}
\newcommand{\eeas}{\end{eqnarray*}}

\documentstyle[titlepage,12pt]{article}
\begin{document}
\begin{tabbing}
\` FIT-HE-97-21\\
\` Aug. 1997\\
\` hep-th/xxx
\end{tabbing}
\addtolength{\baselineskip}{0.2\baselineskip}
\begin{center}
\vspace{36pt}
  {\large \bf Solitonic d-5 Brane and Vortex Defects on the World-Sheet}
\end{center}
\vspace{36pt}
\begin{center}
{\bf Kazuo Ghoroku}\footnote{e-mail: gouroku@dontaku.fit.ac.jp}
\end{center}
\vspace{2pt}
\begin{center}
Department of Physics, Fukuoka Institute of Technology,Wajiro, Higashiku, 
Fukuoka 811-02, Japan
\end{center}
\vspace{36pt}
\begin{center}
{\bf abstract}
\end{center}
We examine the behavior of the vortex defects on the world-sheet
when a solitonic d-5 brane is existing in the 
d-dimensional target space, and 
it is seen that the vortices of any charge should be dissociated 
to the free-gas phase when they approach to the brane. The meanings
of this fact are addressed through the analysis of the renormalization
group equations on the world sheet.
\newpage

\section{Introduction}
The rapid developments in the recent string theories have made clear
the importance of the (D)p-branes in understanding 
the non-perturbative aspect of string- and field-theories. So it is
meaningful to examine the properties of p-branes from various point
of views. 

There is an approach to this subject from the world-sheet action of the
string based on the principle of the conformal invariance.
The world sheet action should be corrected when the branes are in the 
target space. 
As an example, the recoil effect of the D-brane coming from the scattering
of the string and the D-brane has been presented on the world sheet 
action in terms of the appropriate operators ~\cite{fis,kog}.
Another kind of effect
of the D-brane on the world sheet has been proposed
\cite{ellis} in terms of the deformed sine-Gordon action. In this proposal,
the sine-Gordon term is introduced as the representation
of the appearance and disappearance of the virtual D-branes.

Here we examine the problem of the solitonic p-brane in a 
(noncritical) d-dimensional target-space through the world-sheet action,
in which the sine-Gordon term is introduced as the representation of the
vortex defects on the sheet. And this term is expected to be
a kind of medium which provides
the dynamical information of the background including
the p-brane. As for the p-brane configuration,
we consider the d-5 brane, which is an extended form of the 5-brane 
solution obtained in the $10d$ supergravity \cite{str}. And 
this solution plays an important role in 
understanding the duality of the strings in ten-dimension.

The sine-Gordon term, which is obtained on the sphere, has
been used in ~\cite{ellis}, but this term would
receive a modification from the requirement of the conformal invariance
~\cite{amb,sch} of the world-sheet action in our approach. 
This modification reflects
how the brane configuration in the target space
and the sine-Gordon term on the world sheet are affected each other.
Our purpose in this paper is to see what kind of the implications 
are obtained from the modified sine-Gordon term.
This analysis would shed some light on the problem of the vacuum
configuration of the target space. 

Since we are considering in the noncritical target-space dimension, 
the world-sheet action contains the Liouville
field or the conformal mode of two dimensional gravity quantized in the
conformal gauge.
In the previous works
~\cite{ellis,tho}, the sine-Gordon term is written by
the Liouville field. However, there is no reason to take this assignment
in our case. In fact,
any bosonic field on the world-sheet is useful to represents 
the sine-Gordon potential, and it could lead to the
the desirable vortex free-energy. Here the sine-Gordon field is not
assigned to the Liouville field but to some other scalar field on the 
world-sheet.
The Liouville field is insteadly
used here to represent a physical scale in the 2d curved surface
\cite{amb} in order to derive the renormalization group equations
of the parameters of the world-sheet action.
When the sine-Gordon field is assigned to one of the coordinates
transverse to the d-5 brane, we can show that the sine-Gordon potential
becomes relevant near the brane even if how large the vortex charge is.
Namely, the vortices are in the gaseous phase. In other words, 
any vortex-pair strongly
bounded through a large vortex charge dissociates to the free gas
due to the d-5 brane. 

In the next section, the solitonic d-5 brane is derived
from the low-energy effective target-space action.
Then, vortices are introduced in $\S$3 in terms of the sine-Gordon action.
This is done such as the theory is conformaly invariant. 
In $\S$4, the renormalization
group analysis of the parameters are studied, and the summary and
discussions are
given in the final section.

\section{d-5 Solitonic Brane}

In d=10, N=1 supergravity, 5-brane solitonic
solution is given in \cite{str}. Considering this solution as a prototype,
we extend it to the case of the non-critical dimension, $d\ne 10$.

Our starting point is the following low energy effective action 
~\cite{coop,das} of d-dimension,
\beq
S_t=
{1 \over 4\pi}\int\,d^dX\sqrt{G} e^{-2\Phi}
            \bigl[ R-4(\nabla\Phi)^2+
                  (\nabla T)^2+v(T)+{1 \over 2\times 4!}H_3^2-\kappa
            \bigr],
\label{eq:n2}
\eeq
\noindent where $H_3$ is the field strength of the $2$-form potential
($B_{MN}$) given in \rf{eq:w8} and 
\beq
v(T)=-2T^2+{1 \over 6}T^3+\cdots \  \ . \  \label{eq:n3}
\eeq
The parameter $\kappa$ is remained
finite since noncritical dimension d is considered. For bosonic case,
it is written as, $\kappa=(25-c)/3$, where $c$ represents the central charge
of the theory. Other differences of this action from the supersymmetric one
in \cite{str} are (i) the absence of the Yang-Mills field and (ii)
the presence of the tachyon field. The first point is not important since
we are considering a simple solution, but the tachyon term is necessary here
since this part is assigned to the sine-Gordon potential in the 
world sheet action, which is introduced in the next section.

The solution of \rf{eq:n2} is obtained by solving the following
equations of motion,
\bea
\nabla^2T-2\nabla\Phi \nabla T&=&{1 \over 2}v'(T),\label{eq:n4} \\
 \nabla^2\Phi-2(\nabla\Phi)^2&=&-{\kappa \over 2}
            +{1 \over 2}v(T)-{1 \over 6}H_3^2, \label{eq:n5} \\
 R_{MN}-2\nabla_M\nabla_N \Phi &=&
      -\nabla_M T\nabla_N T-{1 \over 4}H^M_{KL}H^{NKL}, \label{eq:n6} \\
 \partial_M(\sqrt{G}{\rm e}^{-2\Phi}H^{MNK})&=&0, \label{eq:d4}
\eea
\noindent where the indices $M,N,\cdots$ run over d-dimensions,
$0\sim d-1$.
The d-dimensional coordinates are denoted as $X^M=(x^{\mu}, y^m)$,
where $\mu=0\sim d-5$ and $m=d-4\sim d-1$. Here, $x^{\mu}$
denotes the coordinates of the world volume of d-5 brane, and
$y^m$ are the transverse ones to the brane. Further,
the "time" $x^0$ is identified with the Liouville field or 
the conformal mode ($\rho$) on the world sheet.

We solve the above equations in the Euclidean metric by assuming $T=0$
and taking the following ansatz, 
\bea
ds^2 &=& \delta_{\mu\nu}dx^{\mu}dx^{\nu}+{\rm e}^{2B(y)}\delta_{mn}
           dy^m dy^n, \label{eq:d1} \\
H^{mnl}&=&{2\epsilon^{mnls} \over 
             \sqrt{{\rm det}g_{ij}}}\partial_{s}\Phi(y), \label{eq:d2}
\eea
where $y=\sqrt{\delta_{mn}y^my^n}$, the distance from the brane. And
$g_{ij}$ represents
the metric of the transverse four dimensional space. Tangential components
of $H_3$, whose indices include at least one of the tangential directions 
denoted by $\mu$, are zero. 

Before solving our equations, we remind the supersymmetric 5-brane solution 
for the case of $d=10$ ($\kappa=0$). It has been 
obtained \cite{str} in the following form,
\beq
\Phi(y)=B(y)=\Phi_p, \qquad 
   {\rm exp}(2\Phi_p)={\rm e}^{2\phi_0}+{Q_D \over y^2},
        \label{eq:d3}
\eeq
\noindent under the above ansatzs \rf{eq:d1}, \rf{eq:d2}.
Here, $\phi_0$ is a constant and $Q_D$ denotes the
"magnetic" charge, which is supported by $H_3$,
of the 5-brane. 

Here we must find a solution for d$\neq$10 and $\kappa\neq0$. 
For the case of $\kappa=0$, the extended solution
to the diverse dimension ($d\neq 10$) has been given in \cite{duff}, and 
the solution is written by one harmonic function similar to \rf{eq:d3}.
However, in the case of $\kappa\neq 0$, the asymptotic 
form of the d-5 brane solution should take 
the linear dilaton vacuum.
So we must modify at least the dilaton part in the solutions obtained 
in ~\cite{str,duff}. The simplest
solution is obtained by changing the configuration of
the flat inner space
$x^{\mu}$, $\mu=0\sim d-5$, of the brane to the linear dilaton form.
Then our solution is obtained in the following form,

\beq
\Phi=\Phi_0+\Phi_p(y), \qquad \Phi_0={1 \over 2}Q\rho,
           \qquad B(y)=\Phi_p(y),
        \label{eq:d5}
\eeq
\noindent where $\Phi_p(y)$ is the one given in \rf{eq:d3},
and $Q=\sqrt{\kappa}$. 

Here we give some comments on this solution. (i) It may be possible
to consider a more general form for $\Phi_0$ in terms of the constants 
$c_{\mu}$
as follows, $\Phi_0=\Sigma c_{\mu}x^{\mu}/2$ with $\Sigma c_{\mu}^2=\kappa$.
However, we do not take this general form here since we are considering
a theory, in which the dilaton should have its 
asymptotic form of \rf{eq:d5}. This is naturally derived from the
quantization of the string theory at noncritical dimension. 
(ii ) $B(y)$ is different from $\Phi$ by $\Phi_0$. Then this solution would
not satisfy the requirement of the supersymmetry, which demmands 
that the solution
must be written by one common function. So the d-5 brane given here would not
keep a supersymmetry. (iii) Due to the linear dilaton part $\Phi_0$,
it seems impossible to obtain a "fundamental" string 
soliton-solution, which can be
obtained from the combined action of 
$S_t$ (\rf{eq:n2}) and the two dimensional 
source action ($S_2$) embedded in the d-dimension. 
For $\kappa=0$, this solution can be found even if $d\neq 10$ ~\cite{duff,horo}.

\section{Vortices and Monopoles}

Next, we introduce the vortices on the world-sheet from the viewpoint
of the path-integral formulation of the world-sheet action. 
The action, which is responsible for the target space action \rf{eq:n2},
can be written in the form of the following non-linear sigma model,
\bea
    S&=& \frac{1}{ 4 \pi  } \int d^2 z \sqrt{\hat g}[{1 \over 2}(G_{MN}(X) 
           + B_{MN}(X))
 ({\hat g}^{\a\b} + i{\epsilon^{\a\b} \over \sqrt{{\hat g}}}) 
              \partial_{\a} X^M \partial_{\b} X^N 
               \nonumber \\
   & & \qquad \qquad   {}  +{\hat R}\Phi(X)+T(X)] \, \, , \label{eq:w8} 
\eea
\noindent where $X^M$ are assigned as in the previous section,
$X^M=(x^{\mu}, y^m)$.
The theory on the sheet 
is quantized in
the conformal gauge, $g_{\mu\nu}={\rm e}^{2\rho}{\hat g}_{\mu\nu}$,
and the conformal mode ($\rho$), which is aliving through the quantum measure,
is assigned as the "time", $x^0=\rho$.

The vortex configuration on the world-sheet can be introduced
as a topological defect \cite{zin} through a scalar field, say $X_v(z)$. 
The effective action of such vortices can be represented
by the sine-Gordon potential, $\cos (pX_v)$, where $p$ denotes
the vorticity or the vortex charge. Then the gas of vortices, whose total
charge is zero, is expressed by the following lagrangian,
\beq
  {1 \over 2}(\partial X_v)^2+\lambda \cos(pX_v), \label{eq:vor}
\eeq
for the flat surface. Here $\lambda$ is a parameter for the vortex gas
and it will be discussed in the following.

From the viewpoint of the conformal field theory, the same partition
function with the one for the vortex gas can be derived \cite{ovr}
in terms of the following potential,
$$  \lambda \cos (q[X(z)- X({\bar z})]), $$
where we notice that the usual scalar field has the plus combination,
$X(z,{\bar z})=X(z)+X({\bar z})$. The corresponding configuration
to the minus combination is called as the monopole \cite{ovr} which is dual 
to the vortex in the
sense that the vortex charge $p$ and the monopole charge $q$ satisfies
the Dirac quantization condition, $pq=2\pi n$, where $n$ is some integer.
This monopole defect is also considered in \cite{ellis}
as the representation of the apperance and disappearance of virtual D-branes. 
We will briefly comment on this monopole below.

First, we consider the case of vortex which
can be included in the above action
\rf{eq:w8} by making an assignment of $X_v$ to the one of $X^M$
in \rf{eq:w8} and adding the sine-Gordon potential of \rf{eq:vor} which
is written by the assigned coordinate field. But it should be modified
due to the quantization of the 2d surface, so our first task is to find
the modified sine-Gordon potential.
Since this newly added potential can be assigned to the tachyon part,
$T(X)$ of \rf{eq:w8}, then its modified form can be
obtained by solving the equation \rf{eq:n4}.
This is corresponding to obtaining a conformal invariant action under the
condition that the added sine-Gordon term is a perturbation to a
conformal invariant vacuum, which is described here by the d-5 brane.
According to the procedure of \cite{amb}, we assume that the modification
for the sine-Gordon potential can be expressed by the dressed factor,
which is given as follows,
\beq
  T_v=\lambda {\rm e}^{\gamma\rho}\cos(pX), \label{eq:v3}
\eeq
where the original factor ${\rm e}^{2\rho}$
is replaced by ${\rm e}^{\gamma\rho}$. 
Then from the equation \rf{eq:n4}, we obtain the value of $\gamma$
by assuming $\lambda$ being small
and linearizing the equation with respect to $T=T_v$.

In solving \rf{eq:n4}, we should notice the following points; (i) 
The assignments of $X_v$ to the one of $X^M$
are separated into the following two groups,
(a) the one of $x^{\mu}$ $(\mu\ne 0)$ (the inner space
of the d-5 brane) and (b) the one of $y^m$ (the transverse space). 
Although we need a modification of the potential
in both cases, the second case (b) is more
intersting as shown below.
(ii) Secondly, the d-5 brane background given above must
be considered in writing explicitly \rf{eq:n4}.

The linearized form of \rf{eq:n4} with respect to $T$ is written as
\beq
(G^{MN}\partial_M\partial_N-Q\partial_0+2)T=0, \label{eq:v4}
\eeq
where we have used \rf{eq:d1} and \rf{eq:d5}, and the term like 
$\partial B\partial T$ has cancelled out due to the characteristic form of
the d-5 brane. 

First, we consider the case of (a) where $X$ being assigned to
$x^{\mu}$. Substituting \rf{eq:v3} to \rf{eq:v4}, then we obtain
\beq
 \gamma={1 \over 2}(Q-\sqrt{Q^2+4p^2-8}), \label{eq:v5}
\eeq
which is the same one obtained in the linear dilaton vacuum previously
\cite{amb} for the sine-Gordon model coupled to the 2d gravity.
At this order of the approximation, the d-5 brane does not affect on
$\gamma$. Then the Kosteritz-Thouless
(KT) transition point, $p^2=2$ which is given in \cite{amb}, is not changed.

Next, consider the second case (b), where $X$ is assigned to one of $y^m$. 
In this case,
the coefficient of the differential operator depends on $y$, so we
solve the equation by assuming the constancy of $y$. Namely, we
consider the problem on a special surface in the target d-dimensional
space, where the distance ($y$) from the d-5 brane is fixed.
Then we obtain the following solution,
\beq
 \gamma={1 \over 2}(Q-\sqrt{Q^2+4{\tilde p}^2-8}), \label{eq:v55}
\eeq
where
\beq
 {\tilde p}^2={\rm e}^{-2\Phi_p(y)}p^2, \label{eq:v6}
\eeq
and $\Phi_p(y)$ is given in \rf{eq:d3}. In this case, the critical 
vortex-charge varies with $y$, and it is given by
\beq
 p^2_{\rm cr}=2{\rm e}^{2\Phi_p(y)}
             =2(1+{Q_D \over y^2}), \label{eq:v7}
\eeq
which approaches to infinity when $y$ goes to zero, just on the d-5 brane.
This fact implies that all of the vortices of any charge
are in the plasma phase near the
d-5 brane. Namely, the vortex and anti-vortex
pair dissociate to the free gas even if how large the vortex charge is.

Nextly, we consider the case of monopoles. Since the monopole charge ($q$)
and the vortex charge ($p$) should satisfy the Dirac condition, $pq=2\pi n$ 
where $n$ is the integer, the monopole is expected to be
in dipole phase if the vortex is in the gaseous phase. This expectation is
true for the flat background. We examine this point 
in the case of the presence of the d-5 brane in the background. 
Then, we solve \rf{eq:v4} by
replacing $T$ by
\beq
  T_m=\lambda' {\rm e}^{\gamma'\rho}
          \cos(p[X(z)-X({\bar z})]). \label{eq:v8}
\eeq
In this case, we must define the operation of $\partial^2$ on $T_m$
in \rf{eq:v3}. Since $\partial^2$ is the laplacian in the flat space,
it might be replaced by the Virasoro operator, $L_0+{\bar L}_0$, defined
for a free scalar field $X$. According to the usual operator formalism,
\bea
   X(z)&=&{x \over 2} -ik\ln z+i\sum_{m\neq 0}{1 \over m}a_mz^{-m}, \label{eq:v9} \\
   X({\bar z})&=&{x \over 2} -ik\ln{\bar z}+
              i\sum_{m\neq 0}{1 \over m}a_m{\bar z}^{-m}, \label{eq:v10} \\
   L_0&=&{\bar L}_0= \sum_{m=1}^\infty :a_ma_{-m}: \, , \label{eq:v11}
\eea
where $x$ is the center of mass of $X$, and $k$ can be taken as zero 
except for the Liouville field. 
Since we do not assign $X$ to the Liouville field
here, $k$ is taken zero. Then we can solve \rf{eq:v4}, and we obtain 
$\gamma'$ in the same form with 
\rf{eq:v5} and \rf{eq:v55} for the two kinds of assignments respectively. 
This result implies
that the dual phase relation of monopole and the vortex is broken near the
brane since both critical charges become infinite near the d-5 brane and 
both defects would be in the plasma phase near the d-5 brane. 
Then we can say that the d-5 brane
dissociate all the topological defects, which are in a dipole
phase, into the free-gas phase. 

\section{Renormalization group Analysis}

In order to understand the above result more deeply,
we examine here the behaviors
of the parameters, $\lambda$ and $p$,
of the vortex through the renormalization group analysis.
The renormalization group equations of these parameters are obtained 
according to the method in \cite{amb,sch}. 
It is as follows.
First, the effective action on the world sheet is separated into the
conformally exact part and a small perturbation which is characterized
by a small parameter (here $\lambda$). To restore the conformal invariance,
which is broken by the small perturbation, the first exact solution must be
modified order by order.
Namely the new exact solution can be expanded in the power
series of $\lambda$. After getting the effective action
at some order of $\lambda$, the renormalization group
equations are obtained by shifting the conformal field, $\rho$, by a constant
in the world sheet action and absorbing this shift in the parameters.
For the first order approximation,
the effective action can be obtained by solving the equations \rf{eq:n4}
$\sim$ \rf{eq:d4}.

Strictly speaking,
the configuration of d-5 brane given here is not an exact one since
the starting action \rf{eq:n2} is an approximate target space action
where higher derivative terms and the massive modes
are abbreviated. As an exact solution, we consider here the 
linear dilaton vacuum, $G_{MN}=\delta_{MN}, \Phi=\Phi_0$
and others are zero. So the d-5 brane solution is expanded around this
vacuum configuration in powers of $Q_D/y^2$, which is assumed small, and 
further add the sine-Gordon term as a small perturbation. Therefore,
there are two small parameters in this case, $\lambda$ and $Q_D/y^2$.
For the sake of the brevity, we consider the case of 
$Q_D/y^2\sim \lambda^2$ hereafter.

First, we consider the case of the assignment, $X_v=x^1$.
Far away from the brane, where 
the relation $Q_D/y^2\sim\lambda^2$ is satisfied, we can expand
$G_{\mu\nu}$, $\Phi$, $H$ and $T$ as follows,
\bea
G_{MN}&=& \delta_{MN}
                +\lambda^2h_{MN}+{Q_D \over y^2}\delta_M^m\delta_N^m
                         + \cdots , \label{11} \\
\Phi&=&\Phi^{(0)}+\lambda^2\Phi^{(2)}+{Q_D \over 2y^2}+
                     \cdots, \label{12} \\
T&=&\lambda(T^{(0)} +\lambda T^{(1)}+\cdots),   \label{13} \\
H^{mnl}&=& -\epsilon^{mnls}{y^s \over y^4}Q_D+\cdots, \label{133}
\eea
where $\cdots$ denotes the higher order terms. Since $H$ is the order
of $1/y^3$, this field can be neglected hereafter in solving the equations.
Since $T^2$ is the lowest power in the
eqs.\rf{eq:n5} and \rf{eq:n6}, it follows that the lowest order
corrections to $G_{\mu\nu}$ and $\Phi$ are of the order $O(\lambda^2)$.
Then, we make an anzatz,
\beq
       h_{MN}=\delta_M^1\delta_N^1h(\rho), \label{14}
\eeq
as in \cite{amb}. The reason of this setting is that the lowest 
order correction 
should appear as the renormalization of the field $x^1$ because of its
self-interaction through the sine-Gordon potential. Then,
the equations of $O(\lambda^2)$ are obtained as follows,
\bea
(\partial^2-2Q\partial_0)\Phi^{(2)}+
{Q \over 4}\partial_0 h&=& -T_0^2,
                                          \hspace{.5cm}\label{f1}\\
2\partial_0^2\Phi^{(2)}-
{1 \over 2}\partial_0^2 h&=& (\partial_0T_0)^2,
                                          \hspace{.5cm}\label{g00}\\
    2\partial_1\partial_0\Phi^{(2)}
   &=&\partial_0T_0\partial_1T_0 ,
                                          \hspace{.5cm}\label{g01}\\
{1 \over 2}(\partial_0^2-Q\partial_0)h
                    -2\partial_1^2\Phi^{(2)}
            &=&-(\partial_1T_0)^2,
                                          \hspace{.5cm} \label{g11}\\
\partial_0\partial_i\Phi^{(2)}=
\partial_1\partial_i\Phi^{(2)}&=&
\partial_i\partial_j\Phi^{(2)}=0,
                                          \hspace{.5cm} \label{gij}\\
(\partial^2-Q\partial_0+2)T_1
            &=&{1 \over 4}T_0^2,
                                          \hspace{.5cm} \label{t1}
\eea
where the higher order terms like $1/y^3$ are neglected.
As a result, these equations are equivalent to the one obtained
under the assumption of $Q_D/y^2<<\lambda^2$ since the terms depending
on $Q_D$ cancel or are neglected as the higher order terms.

Rescaling $T$ by factor four, these equations are solved
near $\gamma\sim 0$, and we obtain
\beq
\Phi^{(2)}={1 \over 256}\cos(2p\phi)+{1 \over 32Q}\rho+O(\gamma),
\label{22}
\eeq
\beq
h ={p^2 \over 16Q}\rho+O(\gamma), \label{23}
\eeq
\beq
T_1={1 \over 32}(1-\cos 2p\phi)+O(\gamma). \label{27}
\eeq
As a result, the effective action is obtained up to $O(\lambda^2)$ 
near $\gamma=0$ where $p=\sqrt{2}$. Then, the renormalization group 
equations are obtained as mentioned above through a shift 
$\rho \to \rho +2dl/\alpha$. Here, $dl$ denotes a small scale of the length
and $\alpha$ represents the dressed factor for the cosmological constant.
And, $\alpha$ is determined by solving equation \rf{eq:v4} with 
$T=\mu^2{\rm exp}(\alpha\rho)$. Denoting as 
$p = \sqrt{2}+\epsilon$, we obtain
\beq
  \dot{\lambda}={4\sqrt{2} \over \alpha Q}\epsilon\lambda, \qquad
   \dot{\epsilon}={\sqrt{2}p^2 \over 16\alpha Q}\lambda^2, \label{33}
\eeq
where $\dot{a}$ means the derivative, $\dot{a}=-da/d\ln{l}$.
From these results, we can see that the sine-Gordon term is irrelevant for 
$\epsilon>0$ where the vortex is in dipole phase. On the other hand,
it becomes relevant in the gaseous phase where 
$\epsilon <0$. And $\lambda$ becomes large in the
infrared limit and the coordinate of the target space
assigned to $X_v$ would be fixed at some value of the periodic copies.

For the case of the assignment, $X_v=y^m$ ($m=d-4$), the 
ansatz \rf{14} for the perturbation of the metric is replaced by,
\beq
       h_{MN}=\delta_M^{d-4}\delta_N^{d-4}h(\rho), \label{34}
\eeq
and other fields are expanded in the same way with the former case. 
The resultant equations of motion
and the solutions are equivalent to the above case
except for the fact that the role of the suffix $1$ is replaced by $d-4$
in this case, and we obtain the same renormalization 
group equations of $\lambda$ and 
$\epsilon$ with the ones given in \rf{33}. 

As mentioned in the previous section, the vortices of any charge, which
is large enough to bind them to the dipole pairs, 
are in the gaseous phase ($\epsilon <0$) near the brane
when we assigned as $X_v=y^m$. As a result,
the sine-Gordon term becomes relevant there in this case, and the coordinate
$y^m$ might be fixed at some value.
This phenomenon seems to imply that the effective central charge 
of the noncritical string would be suppressed by the d-5 brane,
since one of $y^m$ does not contribute to the central charge. 
In this sense, the
brane could open a new possibility of the noncritical string theory in the
higher dimension ($d>2$).

Another aspect of this result is a possibility to interpret this phenomenon
as a signal of a
dynamical compactification since one of $y^i$ could be restricted
to a finite region due to the sine-Gordon potential.
If this idea is possible, it is
intersting to consider an extended form of the sine-Gordon potential such
as
\beq
  T_v=\lambda {\rm e}^{\gamma\rho}
             \sum_{i=d-4}^{d-1}\cos(p_iy^i), \label{eq:dis1}
\eeq
and to see in what circumstances this potential becomes relevant or not.
It is easy to perform the same analysis given above with this potential,
but we can not obtain a simple result of the renormalization group
equations since new counter terms are necessary 
even in the first order of the approximation in this case. Then, it seems
to be necessary to extend the approximation of the analysis to more 
higher orders in order to arrive at a meaningful result. This is out
of our scope here.

\section{Summary and Discussions}

The solitonic d-5 brane is considered in the 
d-dimensional target space as
an extension of the 5-brane in the 10d supergravity.
From the standpoint of noncritical string, 
the influence of the d-5 brane on the world-sheet action
is examined through the renormalization group analysis
on the world-sheet, where the sine-Gordon term is introduced 
to represent the vortex defects from the path-integral formalism.

Our result implies
that the allowed region of one of the coordinate $y^i$ (the transvese
direction of the d-5 brane) 
might be restricted to the region 
around the minimum of the sine-Gordon potential
when $y^i$ is assigned to $X_v$ (the vortex field). 
This means that at least one scalar field on the world-sheet becomes
effectively a constant and one central charge vanishes as a result.
Then the d-5 branes might be considered as
a key ingredient in obtaining a new noncritical string.

On the other hand,
this phenomenon might be related to the dimensional compactification
of the direction represented by $y^i$.
If this idea is meaningful, 
it is intersting to extend the form of the sine-Gordon potential such that
it includes many coodinates. Analysis in this direction will be given
in the near future.

Finally we comment on the possibility of the sine-Gordon term as a realistic,
condensed form of the tachyon part of a noncritical string as suggested
from a different origin \cite{gho1}, where the condensed tachyon has
a different form. In this case,
the string fields propagate in the potential which is periodic 
in some direction. Then the spectrum of the field shows a band structure
as we can see in the case of an electron within a crystal. 
It is interesting if we can 
observe such a evidense in some field's spectrum in our macroscopic world.

\vspace{2cm}
{\bf Acknoledgment}: The author thanks to the members of the high-energy
group of Kyushu University for useful discussions and comments.


\end{document}